\documentclass{article}      
\usepackage{graphicx}
\usepackage{epstopdf}
\title{The Hypercentral Constituent Quark Model and its symmetry}  

\author{M.M. Giannini, E. Santopinto\\
Dipartimento di Fisica dell'Universit\`a di Genova\\
and \\
 I.N.F.N.
Sezione di Genova, Italy
}

\date{}
\begin{document}

\maketitle  

\begin{abstract}
The hypercentral CQM, which is inspired by
Lattice QCD calculations for quark-antiquark potentials, is presented, stressing its
underlying symmetry. Its results for the spectrum, the helicity amplitudes and the
elastic form factors are briefly reported. In the latter case the model has allowed
to show, for the first time in the framework of a quark model, that relativistic
effects are responsible for a deviation from the usually accepted dipole behaviour,
in agreement with recent data taken at the Jefferson Lab.
\end{abstract}

\section{Constituent Quark Models}

Looking at the baryon spectrum, one notices that the best known resonances
(4* and 3*, according to the PDG classification) can be neatly arranged in
$SU(6)$-multiplets, containing states only partially degenerate. This means
that any CQM must provide a good description of the average values of the
energies, by means of a
$SU(6)$-independent interaction, that is spin-flavour independent, to which a $SU(6)$-breaking
term is added, in order to reproduce the splittings within the various
multiplets.

The various CQMs differ in the treatment of both the $SU(6)$-invariant and
the $SU(6)$-breaking interaction. 

In order to construct the $SU(6)$-configurations it is sufficient to
determine the space part of the wave function, since the spin-flavour parts are standard. 
The relative motion of the three quarks is described by the Jacobi coordinates
\begin{equation}
\vec{\rho}~=~ \frac{1}{\sqrt{2}}
(\vec{r}_1 - \vec{r}_2) ~,\nonumber\\
\\
~~~~\vec{\lambda}~=~\frac{1}{\sqrt{6}}
(\vec{r}_1 + \vec{r}_2 - 2\vec{r}_3) ~.\nonumber
\end{equation}

and therefore here are six space degrees of freedom. This is the starting 
point of the algebraic approach \cite{bil}, which introduces 
$u(7)$ as the spectrum generating algebra and the totally symmetric
representation of $u(7)$ as the corresponding space of the three-quark
states. Moreover, the $u(7)$-algebra admits at least the following
two subalgebra chains:

\begin{equation}\label{eq:a1}
\begin{array}{ccccccc}
& &  U(6) & & & & \hspace{2cm} (I) \\
& \nearrow &  & \searrow & & &  \\
 U(7) & & & & SO(6) & , & \\
& \searrow &  & \nearrow & & & \\
& &  SO(7) & & & & \hspace{2cm} (II) \\
\end{array}
\end{equation}
The first chain corresponds to a spherical oscillator (h.o.) in six dimensions,  while the second
chain to a $SO(7)$ dynamical symmetry, as for the hyperCoulomb potential (hC).

In the $u(7)$ model developed by Iachello, Bijker and Leviatan \cite{bil}, the baryon mass operator is
written in
terms of the normal vibrations of a Y-shaped symmetric top, to which a
rotation band is superimposed:
\begin{equation}\label{u7}
M^{2}_{vibr}÷=÷N÷[k_1÷n_u÷+÷k_2÷(n_v÷+÷n_w)]N÷+÷\alpha L÷+÷M^2_0
\label{eq:u7}
\end{equation}
where $n_u$, $n_v$ and $n_w$ are the vibration quantum numbers. To the
operator of Eq.÷(\ref{eq:u7}) a G\"{u}rsey-Radicati term is added:
\begin{equation}\label{eq:gr}
M^2_{sf}÷=÷a÷[C_2(SU_{sf}(6))÷-÷45]÷+÷b÷[C_2(SU_{f}(3))÷-÷9]÷+÷
c÷[C_2(SU_{s}(2))÷-÷\frac{3}{4}]
\end{equation}
where the quantities $C_2$ are the quadratic Casimir operators of the groups
indicated within parentheses. The term in Eq.÷\ref{eq:gr} is $SU(6)$
violating, since it introduces a dependence of the energy on the spin and
isospin, but it does not mix the $SU(6)$-configurations. The $u(7)$ model
leads to a good description of the spectrum and of other baryon
properties.

In order to follow one of the two chains of Eq. (\ref{eq:a1}), one has to
introduce an explicit quark interaction. To this end it is convenient to
substitute the Jacobi coordinates with the hyperspherical ones, which
keep the four angles $\Omega_{\rho}$ and $\Omega_{\lambda}$, but replace
$\rho$ and $\lambda$ with the hyperradius $x$ and the hyperangle $\xi$
\begin{equation}
x=\sqrt{{\vec{\rho}}^2+{\vec{\lambda}}^2} ~~,~~ \quad
\xi=arctg(\frac{{\rho}}{{\lambda}}).
\end{equation}
Using these coordinates, the kinetic term in
the three-body Schr\"{o}dinger equation can be rewritten as \cite{baf}
\begin{equation}
- \frac{1}{2m} ({\Delta}_{\rho}+{\Delta}_{\lambda})= - \frac{1}{2m}
(\frac{{\partial}^2}{{\partial}x^2}+\frac{5}{x} 
\frac{{\partial}}{{\partial}x}-\frac{L^2({\Omega}_{\rho},
{\Omega}_{\lambda},\xi)}{x^2})~~.~~\label{eq:ke}
\end{equation}
where $L^2(\Omega_{\rho},\Omega_{\lambda},\xi)$ is the quadratic Casimir
operator of $O(6)$; its eigenfunctions are the well known hyperspherical
harmonics \cite{baf}
${Y}_{[{\gamma}]l_{\rho}l_{\lambda}}({\Omega}_{\rho},{\Omega}_{\lambda},\xi)$
 having eigenvalues $\gamma(\gamma+4)$, with
$\gamma=2n+l_{\rho}+l_{\lambda}$ (n is a non negative integer).

In the hypercentral Constituent Quark Model (hCQM) \cite{pl}, the quark
interaction is assumed to depend on the hyperradius $x$ only $V=V_{3q}(x)$. The hyperradius
$x$ depend on the coordinates of all the three quarks and then $V_{3q}(x)$ is a three-body
interaction. Actually, three-body mechanisms are generated by the fundamental
multi-gluon vertices predicted by QCD. On the other hand, flux tube models lead
to Y-shaped three-quark interactions. Furthermore, a two body potential, treated
in the hypercentral approximation \cite{hca}, leads to a x-dependent potential,
since averaging $\sum_{i<j}÷(r_{ij})^n$ over angles and hyperangle one gets
something proportional to $x^n$.

For a hypercentral potential the hyperradial wave function, ${\psi}_{\gamma}(x)$ is factored
out and is a solution of the hypercentral equation
\begin{equation}
[~\frac{{d}^2}{dx^2}+\frac{5}{x}~\frac{d}{dx}-\frac{\gamma(\gamma+4)}{x^2}]
~~{\psi}_{\gamma}(x)
~~=~~-2m~[E-V_{3q}(x)]~~{\psi}_{\gamma}(x)~~~.
\label{eq:rad}
\end{equation}

The Eq. (\ref{eq:rad}) can be solved analytically in two cases. The
first is the six-dimensional harmonic oscillator (h.o.)
\begin{equation}
\sum_{i<j}~\frac{1}{2}~k~(\vec{r_i} - \vec{r_j})^2~=~\frac{3}{2}~k~x^2~=
~V_{h.o}(x)
\end{equation}
and the second one is the hyperCoulomb (hC) potential
\begin{equation}
V_{hyc}(x)= -\frac{\tau}{x}. \label{eq:hyc}
\end{equation}
 It is interesting to observe that the energy
levels of the potential (\ref{eq:hyc}) can be obtained generalizing the
procedure used in the case of the three-dimensional Coulomb problem \cite{br}. In n
dimensions the symmetry group is $O(n)$, with the degeneracy group $O(n+1)$,
implying the existence of a conserved n-dimensional Runge-Lenz vector.
The energy levels can then be written in terms of the eigenvalues of the quadratic
Casimir operator
\begin{equation}
E_{\nu÷\gamma}÷=÷-\frac{m÷\tau^2}{2(C_2(O(n+1))+(\frac{n-1}{2})^2)}
\end{equation}
Comparing the first radially excited state  (with $\nu=1$ and
$\gamma=0$) and the first negative states (with $\nu=1$ and
$\gamma=1$) in the two potentials, one sees that they are perfectly degenerate
in the hC potential while in the h.o. case the negative state is much lower.
Since the observed first radially excited nucleon state (the Roper) is slightly
lower than the negative parity resonances, the hC potential seems to be a better
starting point for studying the baryon spectrum. Furthermore, the h.o. levels are
too degenerate in comparison with the experimental spectrum.

We shall consider in particular three CQMs:

A) The Isgur-Karl model \cite{ik}:
\begin{equation}
V_{3q}÷=÷V_{h.o}(x)÷+÷U÷+÷H_{hyp},
\end{equation}
where the strong degeneracy of the h.o. levels is modified by means of the shifting
two-body potential $U$ and the splitting within the mujltiplets is produced by the
spin dependent hyperfine interaction $H_{hyp}$.

B) The analytical hypercentral model \cite{sig,sig1}
\begin{equation}
V_{3q}÷=÷-\frac{\tau}{x}÷+÷\beta÷x÷+÷A÷e^{-\alpha x}÷\sum_{i<j} {\vec{\sigma}}_i
{\vec{\sigma}}_j÷+÷tensor int,
\end{equation}
the linear confinement is treated as a perturbation and
therefore the problem can be solved analytically.
The perturbation treatment is justified for the lower levels.

C) The hypercentral Constituent Quark Model \cite{pl}
\begin{equation}
V_{3q}÷=÷-\frac{\tau}{x}÷+÷\alpha÷x÷+H_{hyp}\label{eq:hyp}
\end{equation}
because of the confinement term the hyperradial equation (\ref{eq:rad}) has to be
solved numerically. The form of the hCQM Eq. (\ref{eq:hyp}) is supported by recent
Lattice QCD calculations \cite{bali}, which results in a quark-antiquark potential
containing a coulomb-like term plus a linear confinement. Model C) contains only
three free parameters which can be fitted in order to reproduce the experimental
spectrum \cite{pl}. An improved version of the model includes also isospin
dependent terms \cite{iso}, leading to a very good
agreement with the experimental levels, including the correct order and position of
the Roper and the negative parity resonances. 

\section{The electromagnetic excitation of baryon resonances}

The spectrum is well reproduced by many CQMs, therefore in order to distinguish among
them it is necessary to consider other physical quantities of interest, such as the
photocouplings, the helicity amplitudes, the strong decays and the nucleon elastic
form factors.

The helicity amplitudes are defined as 
\begin{equation}\label{eq:helamp}
A_{M}(Q^2) ~=~ \langle B, J', J'_{z}=M | H_{em}^t| N, J~=~
\frac{1}{2}, J_{z}= M-1\rangle\\ ~~~~~~ M~=~\frac{1}{2},\frac{3}{2}\
\end{equation}
The transverse transition operator is assumed to be
\begin{equation}\label{eq:htm}
H^t_{em}~=~-~\sum_{i=1}^3~\left[\frac{e_j}{2m_j}~(\vec{p_j} \cdot
\vec{A_j}~+
~\vec{A_j} \cdot \vec{p_j})~+~2 \mu_j~\vec{s_j} \cdot (\vec{\nabla}
\times \vec{A_j})\right]~~,
\end{equation}
where spin-orbit and higher order corrections are neglected
\cite{cko,ki,aie}. In Eq. (\ref{eq:htm}) $~m_j$, $e_j$, $\vec{s_j}$ , $\vec{p_j}$ and
$\mu_j~=~\frac{ge_j}{2m_j}$
denote the mass, the electric charge, the spin, the momentum and the
magnetic moment of the j-th quark, respectively, and
$\vec{A_j}~=~\vec{A_j}(\vec{r_j})$ is the photon field. 

In the case of models B) and C), the parameters fitted to the spectrum are used and
therefore the helicity amplitudes are given by parameter-free calculations, while in
model A) the h.o. constant $\frac{1}{\alpha}\simeq 0.5 fm$ is adjusted in order to
reproduce the amplitude
$A_{3/2}$ for the $D_{13}$ resonance at the photon point \cite{cko,ki}.
The photocouplings calculated with these models (and with other models
as well) follow qualitatively the behaviour of the experimental data
\cite{aie}: there is however generally an underestimate of the
observed strength. The similarity of the results is due to the fact
that the models have basically the same underlying spin-isospin symmetry. 
As for the transition form factors, the results for to negative parity
resonances
\cite{aie2} show that the helicity amplitudes calculated with the h.o.
potential have a
$Q^2$ behaviour completely different from data. The models B) and C) 
reproduce the experimental data for medium-high $Q^2$, showing that the hypercoulomb
interactions apparently leads to
more realistic three quark wavefunctions.
At low $Q^2$ there is often a lack of strength, in agreement to what happens at
the photon point, specially in the case of the $A_{3/2}$ amplitudes. This discrepancy
indicates that some mechanism, important at low $Q^2$, is missing, such as the
quark-antiquark pair production \cite{pl,iac}. 
In some cases, the missing strength is less evident, as for the
$S_{11}$ resonances. In Fig. 1 we show the helicity amplitude calculated
with the hypercentral model
\cite{aie2} for the
$S_{11}(1535)$ state in comparison with the model of ref. \cite{ck} and the data. It should be
reminded that the curve has been published three years earlier than the recent
Jlab data \cite{dytman}. Also the amplitude for the $S_{11}(1650)$ state is well
reproduced \cite{aie2} and this is a sensible test of $SU(6)$ violation, since in absence of
any configuration mixing the amplitude should be exactly zero. All these results are only
slightly modified by the introduction of relativistic corrections \cite{mds}. 

%\newline
%\indent
\begin{figure}[!ht]
\vspace*{-1.5cm}
\begin{center}
\includegraphics[width=9cm]{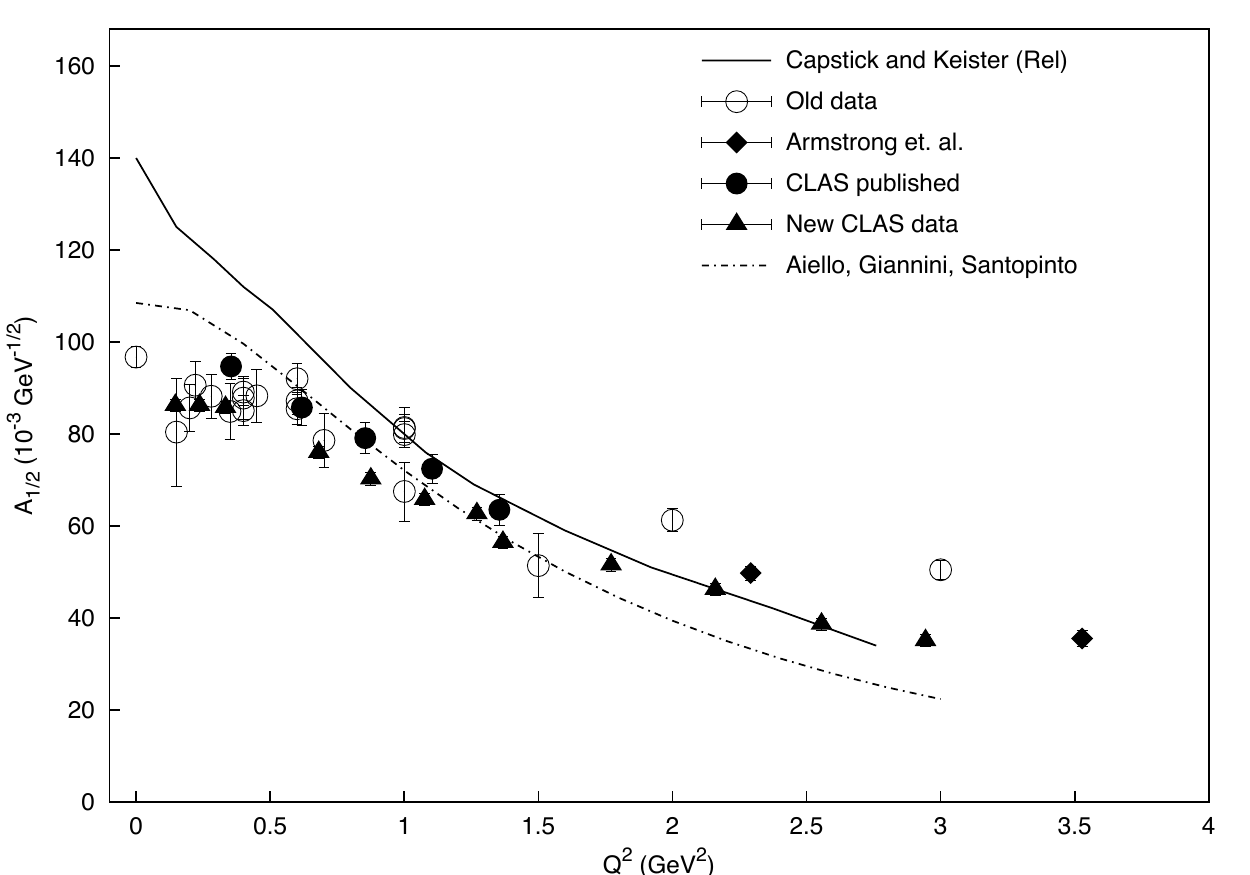}
\vspace*{-0.3cm}
\caption[]{ The helicity amplitude for the $S_{11}(1535)$ state, calculated with the
hypercentral CQM \cite{aie2}, in comparison with the model of ref. \cite{ck} and with data,
including the recent Jlab ones. (Data from a compilation by \cite{burk02}.)}
\end{center}
\end{figure}

\indent
\begin{figure}[!ht]
\vspace*{-0.5cm}
\begin{center}
\includegraphics[width=8cm]{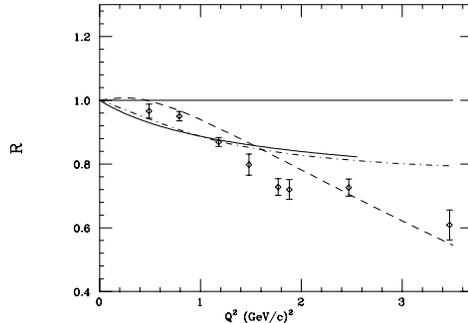}
\vspace*{-1.5cm}
\caption[]{The ratio $R~= {\mu_p}~G_{E}/G_{M}$  calculated with
the hCQM, taking into account the relativistic kinematical corrections
(full line, ref. \cite{rap}). The horizontal full line represents the ratio for
CQM without relativistic corrections. The dashed curve is the
fit of ref. \cite{ijl}, the dot-dashed curve is the dispersion relation fit of
ref. \cite{dd}.
The points are the data of the recent Jlab experiment Ref. \cite{jlab2}}
\end{center}
\end{figure}

\section{The nucleon elastic form factors}
The hypercentral CQM has been used also for the calculation of the elastic nucleon
form factors. To this end some relativistic corrections have been introduced,
boosting the initial and final state to a common Breit frame and expanding the
quark current at the lowest order in the quark momentum \cite{mds2}. In this way it
has been possible to show, for the first time in the framework of quark models
\cite{rap}, that the ratio R between the electric and magnetic form factors of the
proton deviates from the standard dipole value ($\simeq 1$) because of relativistic
effects (see Fig. 2). This behaviour is in agreement with the recent Jlab data
\cite{ped}.

The relativistic corrections to the hCQM have been further improved by introducing
the correct relativistic kinetic energy and using a fully relativistic quark
current. Introducing quark small form factors corresponding, the calculated
nucleon elastic form factors describe the data very well up to $4 GeV^2$ and the
ratio R reaches a value of about 0.6, in better agreement with the new Jlab data
\cite{jlab2}.

\end{document}